\documentclass[11pt,a4paper]{article}
\usepackage{graphicx}
\usepackage{color}
\usepackage{docmute}
\usepackage[left=20mm,right=20mm,top=30mm,bottom=30mm]{geometry}
\usepackage{xparse}
\usepackage{amssymb,amsmath}
\usepackage{mathtools}
\usepackage{physics}
\usepackage{hyperref}
\usepackage{enumerate}
\usepackage{longtable,booktabs}
\usepackage{color, graphicx}
\usepackage{array}
\usepackage{float}
\usepackage{cancel}
\usepackage{bm}
\usepackage{subcaption}
\usepackage{multirow}
\usepackage[normalem]{ulem}

\title{Quantum machine learning interatomic potential:\\
  Application of variational quantum algorithm}
\author{\small Kohei Numata\textsuperscript{1},
  Wataru Mizukami\textsuperscript{2,~3},
  Kosuke Mitarai\textsuperscript{2,~3},
  Keisuke Fujii\textsuperscript{2,~3,~4} \\
  \small and Yutaka Imamura\textsuperscript{1}}
\date{}

\begin{document}
\maketitle
\begin{center}
  \textsuperscript{1} \footnotesize \emph{Innovative Technology Laboratories, AGC Inc.,
  Yokohama, Kanagawa 230-0045, Japan} \\
  \textsuperscript{2} \footnotesize \emph{Graduate School of Engineering Science, The University of Osaka,
    1-3 Machikaneyama, Toyonaka, Osaka 560-8531, Japan} \\
  \textsuperscript{3} \footnotesize \emph{Center for Quantum Information and Quantum
    Biology, The University of Osaka, 1-2 Machikaneyama, Toyonaka, Osaka 560-0043, Japan} \\
  \textsuperscript{4} \footnotesize \emph{RIKEN Center for Quantum Computing (RQC),
    Wako, Saitama 351-0198, Japan} \\
\end{center}

This study applied quantum circuit learning, a commonly used hybrid quantum-classical machine learning algorithm,
to a machine learning interatomic potential (MLIP) for predicting the energies of molecules in molecular datasets.
We retrained the ANI model using the quantum transfer learning architecture [Mari et al., \textit{Quantum}, 4:340, 2020]
and evaluated numerical accuracy with a quantum circuit simulator.
The evaluation confirmed that inserting a quantum circuit into the classical neural network of the MLIP yielded slightly higher accuracy than the fully classical neural network under certain conditions.
In particular, the model incorporating a quantum circuit was more effective
when the pretraining model had room for improvement in accuracy.
These findings may contribute to advancing the application of quantum machine learning for MLIPs.

\section{Introduction}\label{introduction}

The machine learning interatomic potential (MLIP),
which takes the nuclear configuration of a chemical system as input and outputs its total energy,
has attracted significant attention in recent years 
~\cite{behler2007generalized, Bartok2010GAP, Unke2021MLFF, Jacobs2025Guide}.

MLIP enables fast chemical simulations by avoiding direct solutions of the Schrödinger equation
at each point on the potential-energy surface.
Among the various MLIP frameworks, 
neural-network potentials (NNPs), pioneered by Behler and Parrinello \cite{behler2007generalized}, 
have become widely adopted as powerful computational tools.

Their accuracy and transferability have improved dramatically with the introduction of
graph neural networks and $\mathrm{E}(3)$-equivariant message-passing architectures
\cite{batzner20223, Batatia2022mace, Batatia2022Design}.

Combined with large first-principles datasets, these architectural advances
have given rise to universal MLIPs, also referred to as atomistic foundation models, such as
M3GNet \cite{Chen_2022}, CHGNet \cite{Deng2023CHGNet}, MACE-MP-0 \cite{Batatia2025MACEMP0},
GNoME \cite{merchant2023scaling}, Orb \cite{neumann2024orbfastscalableneural, Rhodes2025OrbV3},
and MatterSim \cite{yang2024mattersimdeeplearningatomistic},
which are capable of handling a broad range of chemical species across the periodic table.

Initially, these universal models were trained mainly on inorganic datasets,
but models covering both inorganic and organic systems have recently begun to emerge.
Taking the MACE architecture as an example, the MACE-MH-1 model
accommodates several different levels of theory within a single model using a multi-head structure~\cite{Batatia2025MACEMH1}.
On the other hand, the MACE-Osaka models unify training data computed at different levels of theory
without using multiple output heads~\cite{Shiota2024MACEOsaka24, Kuroda2026MACEOsaka26}.
Another approach, exemplified by UMA, is to train models on multiple datasets
through a mixture of linear experts~\cite{Wood2025UMA}.

Community benchmarks such as Matbench Discovery indicate that state-of-the-art universal models
now attain accuracy approaching that of density functional theory (DFT) for many tasks
while operating several orders of magnitude faster~\cite{Riebesell2025Matbench}.
This success underscores the need for continued advancements in model construction to further enhance performance.

Because interaction energy originates from quantum mechanics, models incorporating quantum features—specifically,
those constructed using quantum computers—may offer additional benefits.
This direction has been partially explored in previous studies.
The earliest attempt involved using a quantum neural network similar to that proposed in Ref. \cite{havlivcek2019supervised}
for this purpose \cite{Kiss2022QuantumNNForceFields}.
Le \textit{et al.} \cite{le2023symmetryinvariantquantummachinelearning} later introduced a quantum encoding of nuclear configurations
that preserves the rotational symmetry of the system, improving upon the earlier approach \cite{Kiss2022QuantumNNForceFields}.
Another method, proposed by Monaco \textit{et al.} \cite{LoMonaco2024QuantumExtremeLearning} employed quantum extreme learning machines,
which are more hardware-efficient \cite{Fujii2017,negoro2018machinelearningcontrollablequantum,Nakajima2019}.
Although these approaches have paved the way for applying quantum machine-learning techniques in this field,
they have so far been limited to small systems, with the largest examples being a water dimer and formamide,
each comprising only six atoms; both remain far from practical molecular systems containing several tens of atoms or more.

More recently, Willow \textit{et al.} proposed a hybrid quantum--classical machine-learning potential
for high-temperature liquid silicon,
where all atom-wise readout functions
are replaced by variational quantum circuits,
while the other blocks 
of the $\mathrm{E}(3)$-equivariant message-passing graph neural network 
are kept classical \cite{Willow2025HQCMLP}.

Here, we propose combining an existing classical MLIP, namely, 
ANAKIN-ME (Accurate NeurAl networK engINe for Molecular Energies),
often abbreviated as ANI \cite{smith2017ani}, 
with a quantum neural network known as quantum circuit learning (QCL) \cite{mitarai2018quantum}.
QCL is one of the variational quantum algorithms \cite{cerezo2021variational}
that performs supervised learning using parameterized quantum circuits.
These circuits can be designed to be relatively shallow,
enabling several experimental demonstrations on actual quantum hardware \cite{havlivcek2019supervised}.
Our strategy follows that of Ref. \cite{mari2020transfer}, specifically replacing
the final layer of the neural network in ANI with a QCL model.

This setting is closely related to the hybrid quantum--classical MLIP recently proposed by Willow \textit{et al.}~\cite{Willow2025HQCMLP},
in the sense that both approaches introduce variational quantum circuits into the readout stage
of an interatomic-potential model rather than directly solving the electronic-structure problem
on a quantum computer~\cite{McArdle2020QuantumComputationalChemistry, Cao2019QuantumChemistryAgeQC, Bauer2020QuantumAlgorithmsChemistry}.

However, the placement and role of the quantum module are different.
Willow \textit{et al.} replaced all atom-wise readout functions attached to the interaction layers
with variational quantum circuits.
In contrast, we first pretrain the ANI model classically, and replace the final layer with a QCL model.

Because ANI produces a highly informative and compressed embedding of the input configuration at its final layer,
we can employ a relatively small QCL model to predict the total energy with minimal effort.

This hybrid approach has two main advantages.
First, since ANI is already trained to output accurate energies,
the QCL-enhanced model can achieve performance comparable to that of the original ANI model.
Second, the compact size of the QCL circuit allows benchmarking our method on classical simulators,
even for larger molecules.
This is in contrast to the earlier studies described above (with the exception of Willow \textit{et al.}),
in which the number of qubits had to scale with
the number of system parameters defining the molecular configuration.
We conducted numerical experiments on this concept and found
that the QCL-enhanced model surpassed the classical neural network in accuracy under specific conditions.
These findings pave the way towards improving MLIP accuracy by incorporating quantum machine-learning techniques.

\section{Methods}\label{methods}
\subsection{Review}\label{review}
Let us first introduce QCL and ANI, the foundational methods employed in this study.
QCL is a hybrid quantum-classical approach that employs parameterized quantum circuits to approximate complex functions.
To make a prediction on an input $\bm{x}$, QCL constructs a model of the following form:
\[
f_\theta(\bm{x}) = \langle \psi(\bm{x}, \bm{\theta}) | O | \psi(\bm{x}, \bm{\theta}) \rangle,
\]
where \(|\psi(\bm{x}, \bm{\theta})\rangle\) is defined as
\[
|\psi(\bm{x}, \bm{\theta})\rangle = U(\bm{\theta})V(\bm{x})\ket{0}^{\otimes n},
\]
and $O$ is a Hermitian operator acting on $n$ qubits.
The parameters $\bm{\theta}$ are optimized using, for example, gradient descent with respect to a chosen cost function.
The use of quantum features, $V(\bm{x})\ket{0}^{\otimes n}$,
has been shown to outperform classical machine-learning methods for certain tasks \cite{Liu2021}.
Moreover, a vector output can be generated within the QCL framework by employing multiple observables $\{O_j\}$.

ANI is a machine-learning potential that employs neural networks trained on quantum-chemical data to predict molecular energies.
It uses atomic environment vectors (AEVs) to encode local chemical environments.
The AEV $\bm{G}_i$ is an $N_G$-dimensional vector defined for each atom $i$ in a system
and represents information about its environment—specifically, which atomic species surround atom $i$ and at what distances.
$N_G$ depends on the number of atomic species considered and on
the hyperparameter settings that define the mapping
$\{\bm{r}_i\} \to \{\bm{G}_i\}$ for a given set of atomic coordinates $\{\bm{r}_i\}$.
For example, in the original ANI model $N_G=768$ \cite{smith2017ani}.
ANI uses a neural network $\mathcal{L}:\mathbb{R}^{N_G}\to\mathbb{R}$
to output the energy of an atom described by its AEV $G_i$.
The total energy of a molecular system characterized by $\{\bm{r}_i\}$ is then given by $\sum_{i} \mathcal{L}(G_i)$.
ANI models achieve near-DFT accuracy while remaining computationally efficient,
making them well-suited for large-scale molecular-dynamics simulations.

\subsection{Insertion of quantum circuits into MLIP}
\begin{figure}
  \centering
  \includegraphics[]{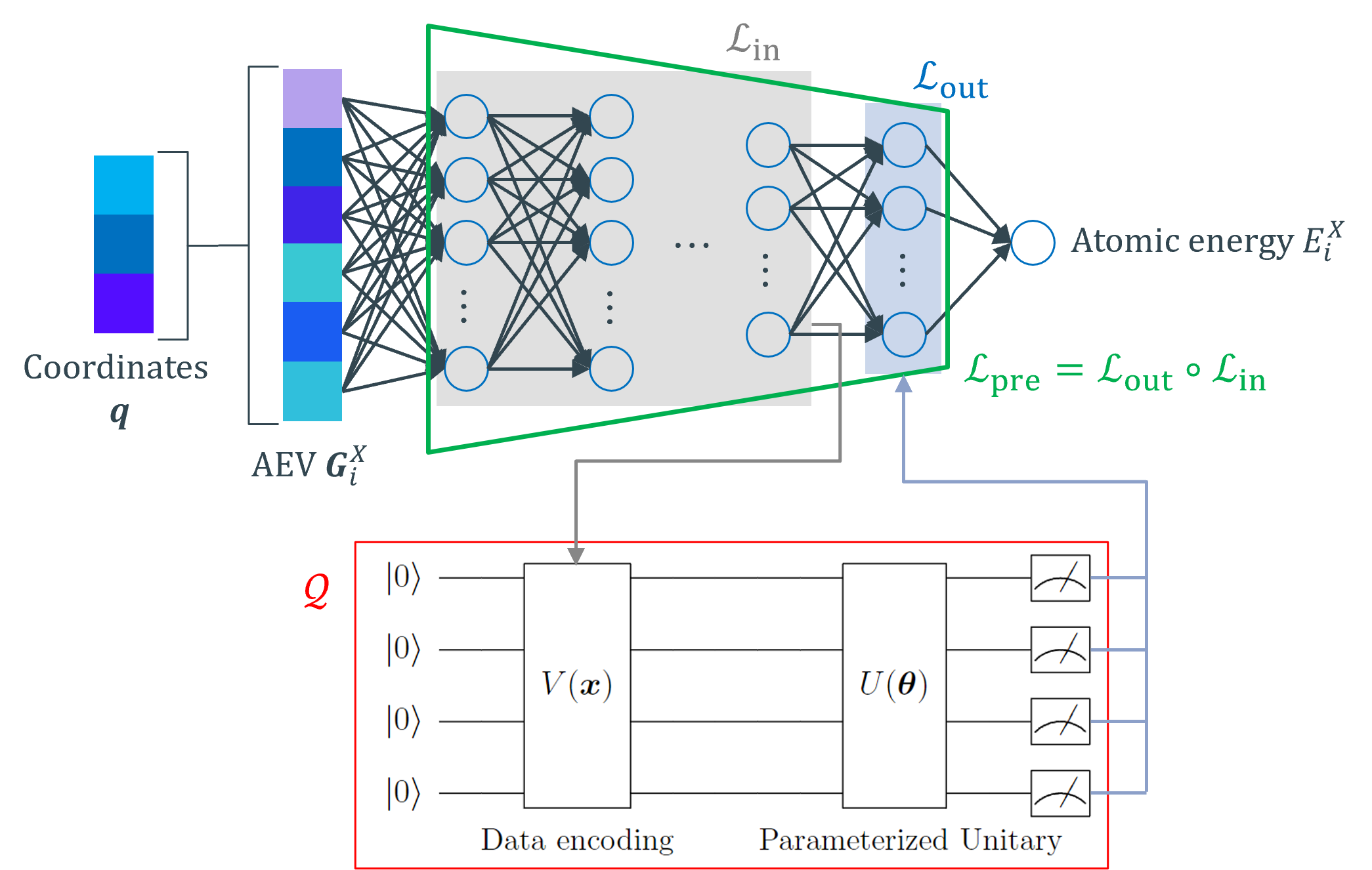}
  \caption{MLIP with a quantum circuit.
    This hybrid architecture is based on the study by Mari et al. \cite{mari2020transfer}.
    The QCL model $\mathcal{Q}$ is inserted between $\mathcal{L}_{\text{out}}$ and $\mathcal{L}_{\text{in}}$,
    which are pretrained within the ANI framework.}
  \label{fig:concept}
\end{figure}

We incorporated a quantum model into the ANI framework following the general methodology proposed in \cite{mari2020transfer}.
Figure \ref{fig:concept} represents an overview of the proposed approach.
Let \(\mathcal{L}_{\text{in}}\) and \(\mathcal{L}_{\text{out}}\) denote fully connected neural-network layers.
The layers consist of $l$ layers and a single layer, respectively.
We assume \(\mathcal{L}_{\text{in}}: \mathbb{R}^{N_G} \to \mathbb{R}^{n_q}\), that is,
the output of $\mathcal{L}_{\text{in}}$ is $n_q \ll N_G$ dimensional.
Correspondingly, $\mathcal{L}_{\text{out}}: \mathbb{R}^{n_q} \to \mathbb{R}$.
First, we trained this network using the ANI framework to optimize the classical parameters before integrating the quantum circuit.
  \begin{align}
      \mathcal{L}_{\text{pre}} = \mathcal{L}_{\text{out}}\circ \mathcal{L}_{\text{in}}
  \end{align}
We refer to this step as pre-training in this work.
Second, we inserted the QCL model $\mathcal{Q}:\mathbb{R}^{n_q}\to \mathbb{R}^{n_q}$
between $\mathcal{L}_{\text{out}}$ and $\mathcal{L}_{\text{in}}$, and trained the model
\begin{align}
    \mathcal{L}_{\rm transferred} = \mathcal{L}_{\rm out} \circ \mathcal{Q} \circ \mathcal{L}_{\rm in}.
\end{align}
Here, we assume that we use the QCL model with multi-dimensional output by using $n_q$ observables $\{O_j\}_{j=1}^{n_q}$.

The strategy adopted here is essentially the same as that proposed in \cite{mari2020transfer}.
However, Mari \textit{et al.} \cite{mari2020transfer} applied their method not to MLIP
but to more conventional machine-learning tasks such as image recognition.
We believe that applying this strategy to MLIP is promising, as
interatomic potentials should ultimately be described using quantum mechanics.
We also designed the quantum circuit with carefully prepared initial values
to verify the improvement achieved by the QCL model over the classical model, as explained in Secs. \ref{circuit} and \ref{inserted-model}.
In what follows, we present numerical simulations to evaluate the performance of this approach.

\section{Numerical performance assessment}
\subsection{Quantum circuit}\label{circuit}

\begin{figure}
  \centering
  \includegraphics[width=4.28426in,height=1.80796in]{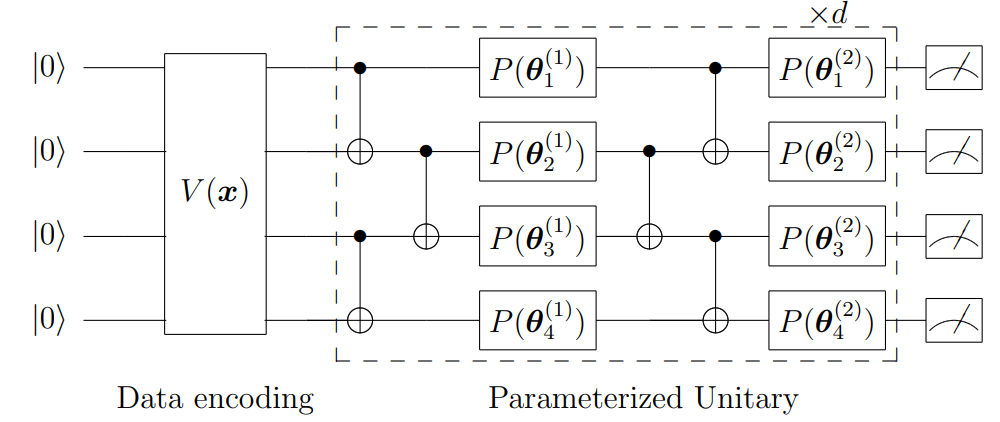}
  \caption{Quantum circuit for the model $\mathcal{Q}$ at $n_{q}=4$.
    \(V(\bm{x})\) is defined in Table \ref{table:circuit-conditions}.
    The circuit enclosed by dotted lines represents the parameterized quantum circuit $U(\bm{\theta})$.}
  \label{fig:circuit}
\end{figure}

We describe the QCL model \(\mathcal{Q}\) used in detail.
For $U(\bm{\theta})$, we used the circuit structure shown in Figure \ref{fig:circuit}.
$P(\bm{\theta})$ denotes a parameterized single-qubit rotation with trainable parameters,
and we tested the combinations of \(V(\bm{x})\) and \(P(\bm{\theta})\) listed in Table \ref{table:circuit-conditions}.
The circuit depth $d$ was set to $d=1$ or 5.
The circuit was designed to reproduce the input value \(\bm{x}\) when all $\bm{\theta}$ were set to zero.
This initialization prevents accuracy degradation due to sensitivity to initial values when the quantum circuit is inserted.
For the observables that define the output of $\mathcal{Q}$,
we used $\{Z_k\}_{k=1}^{n_q}$ where $Z_k$ denotes Pauli-Z operator acting on $k$th qubit.

The numerical assessment was conducted through StateVectorSimulator in Pennylane \cite{bergholm2022pennylaneautomaticdifferentiationhybrid}.
In this study, shot noise and noise from the actual quantum hardware are not taken into account.

\begin{table}[]
  \caption{Quantum circuit setups used in this work. $R_{y}\left( \theta \right)$ and $R_{z}\left( \theta \right)$ denote $y$ and $z$ rotation gates, respectively.}
  \label{table:circuit-conditions}
  \centering
  \newlength{\myheight}
  \setlength{\myheight}{7mm}
  \begin{tabular}{c|ccc}
    \hline
                                      & $V\left( \bm{x} \right)$                                                                                                        & $P\left( \bm{\theta} \right)$                              & \begin{tabular}[c]{@{}c@{}}\# of parameters                                                                                                                                                                                                \\ per 1 depth circuit\end{tabular} \\ \hline
    \parbox[c][\myheight][c]{0mm}{} 1 & $\otimes_{k = 1}^{n_{q}}\left[R_{y}\left( -\arcsin\left( x_k \right) \right)H\right]$                                                     & $R_{y}\left( \theta_{1} \right)$                               & 2$n_{q}$                                                                                                                                                                                                                                                               \\
    \parbox[c][\myheight][c]{0mm}{} 2 & $\otimes_{k = 1}^{n_{q}}\left[R_{z}\left( \arccos\left( x_k \right) \right)R_{y}\left( \arccos\left( x_k \right) \right)\right]$ & $R_{y}\left( \theta_{2} \right)R_{z}\left( \theta_{1} \right)$ & $4n_{q}$                                                                                                                                                                                                                                                               \\ \hline
  \end{tabular}
\end{table}

\subsection{Construction of ANI models}\label{insert-circuit}
The PyTorch-based package TorchANI \cite{gao2020torchani} was used for implementation in this study.
In TorchANI, the default value for $N_G$ is 384, and this value was adopted here.
According to the findings of Smith \textit{et al.} \cite{smith2017ani}, neural networks with three to four hidden layers perform best;
therefore, total layer counts of 3 or 4 (i.e., \(l=2, 3\)) were chosen for this study's pre-training.
In addition to the \(l=2, 3\) configurations, we also examined \(l=1, 4\) to investigate the role of QCL under conditions of insufficient or excessive pretraining.
The numbers of nodes in the intermediate layers, $n_2, n_3, n_4$, are summarized in Table \ref{tab:nodes}.
Note that the input layer was $N_G$ dimensional.
The final output dimension $n_q$ of $\mathcal{L}_{\rm in}$ was varied among 4, 8, and 12.
We used the CELU represented below as an activation function for \(\mathcal{L}_{\text{in}},\ \mathcal{L}_{\text{out}}\):
\begin{equation} \label{eq:celu}
  \text{CELU}\left( x \right) = \max{\left( 0,\ x \right)} + \min{\left(0, \left( \exp\left( 10x \right) - 1 \right)/10\right)}. \\
\end{equation}

During pretraining, we trained the network $\mathcal{L}_{\rm pre} = \mathcal{L}_{\rm out}\circ\mathcal{L}_{\rm in}$
for 100 epochs using mean-squared error as the loss function.
For optimization, we applied AdamW \cite{loshchilov2017decoupled} to the weights
and SGD \cite{robbins1951stochastic} to the biases, with an initial learning rate of 0.001.

The dataset employed in this work was derived from the ANI model library \cite{gao2020torchani}, which contains
the energies of the molecules consisting of H, C, N and O atoms.
60\%, 20\% and 20\% of the dataset were used for training, validation, and test data, respectively.

\subsection{Inserting quantum model}\label{inserted-model}

We divided the original dataset into eight subclasses based on the number of non-hydrogen atoms in each molecule.
More concretely, a dataset $D_n$ contains molecules with $n$ non-hydrogen atoms.
For example, the dataset $D_1$ consists of single non-hydrogen atom molecules such as:
CH\textsubscript{4}, NH\textsubscript{3}, H\textsubscript{2}O.
In this study, we restrict our experiments to $D_1$--$D_4$ (i.e., $n=1,2,3,4$). $D_5$--$D_8$ are excluded from the present work.
This is because $D_5$--$D_8$ involve large target molecules in both size and number,
resulting in high computational cost and making execution impractical.
For each $n=1,2,3,4$, we first divided $D_n$ into training, validation, and test sets with a $6:2:2$ ratio.
For each $n=1,2,3,4$, we then randomly sampled 5,000 training and 2,000 validation data
for training and validating the model $\mathcal{L}_{\rm transferred}$.

During the training of $\mathcal{L}_{\rm transferred}$,
the parameters of $\mathcal{L}_{\rm in}$ and $\mathcal{L}_{\rm out}$ were fixed,
and only the parameters in $\mathcal{Q}$ were optimized.
This procedure ensures performance comparable to the pretrained model while allowing targeted quantum fine-tuning.
For comparison, we also constructed a reference model $\mathcal{L}_{\rm transferred} =\mathcal{L}_{\rm out} \circ \mathcal{L}_{\rm mid} \circ \mathcal{L}_{\rm in}$,
where $\mathcal{L}_{\rm mid}$ is a classical single-layer fully connected neural network with $n_q$ input nodes and $n_q$ output nodes.
These configurations enable a direct performance comparison between the quantum model $\mathcal{Q}$ and its classical neural-network counterpart.
Descriptions of all hyperparameters used in our approach are summarized in Table \ref{table:parameters}.

As in pretraining,
mean-squared error was used as the loss function,
the initial learning rate was 0.001, and
the optimization methods were AdamW and SGD for classical transfer models and
AdamW for quantum transfer models.
The parameters $\bm{\theta}$ of $\mathcal{Q}$ were initialized to zero.
For $\mathcal{L}_{\rm mid}$, the initial weights were set to the identity matrix
so that $\mathcal{L}_{\rm mid}$ reproduced the input value \(\bm{x}\) at initialization.

Model performance was evaluated using the mean root-mean-squared error (RMSE) of the test datasets over five independent trials under each condition.

\begin{table}
  \centering
  \caption{Number of nodes used to construct $\mathcal{L}_{\rm in}$ in each layer \( n_i \) for different values of \( l \), excluding \( n_1 \) which is constant.}
  \label{tab:nodes}
  \begin{tabular}{c|c|c|c}
      \hline
      \( l \) & \( n_2 \) & \( n_3 \) & \( n_4 \) \\
      \hline
      1 & - & - & - \\
      2 & 96 & - & - \\
      3 &
      \( \begin{cases}
          128 & (H) \\
          112 & (C, N, O)
      \end{cases} \)
      & 96 & - \\
      4 &
      \( \begin{cases}
          160 & (H) \\
          144 & (C) \\
          128 & (N, O)
      \end{cases} \)
      & \( \begin{cases}
          128 & (H) \\
          112 & (C, N, O)
      \end{cases} \)
      & 96 \\
      \hline
  \end{tabular}
\end{table}

\renewcommand{\arraystretch}{1.2}
\begin{table}
  \caption{Information on dataset used to train the model $\mathcal{L}_{\rm transferred}$}
  \label{table:dataset}
  \centering
  \begin{tabular}{c|rrr}
    \hline
      & \# of non-hydrogen atoms & \# of total atoms & \# of total data \\ \hline
    $D_1$ & 1                        & 3--5               & 10,799           \\
    $D_2$ & 2                        & 2--8               & 50,961           \\
    $D_3$ & 3                        & 3--11              & 151,199          \\
    $D_4$ & 4                        & 4--14              & 651,935          \\ \hline
  \end{tabular}
\end{table}

\renewcommand{\arraystretch}{1.2}
\begin{table}
  \caption{Descriptions of hyperparameters}
  \label{table:parameters}
  \centering
  \begin{tabular}{c|lr}
    \hline
      Hyperparameter & Description & Value range \\ \hline
    $N_G$ & Input dimension of $\mathcal{L}_{\rm in}$                        & 384           \\
    $l$ & \# of layers in $\mathcal{L}_{\rm in}$                        & 1--4                         \\
    $n_q$ & \# of qubits in $\mathcal{Q}$ / Output dimension of $\mathcal{L}_{\rm in}$   & 4, 8, 12 \\
    $d$ & Circuit depth                        & 1, 5                  \\
    $n$ & Dataset No. / \# of non-hydrogen atoms & 1--4 \\ \hline
  \end{tabular}
\end{table}


\subsection{Numerical assessment for ANI dataset}\label{results_ani}

\subsubsection{Dependence on $n_q$}

Figure \ref{fig:rmse-nq} shows RMSE for the test data
of $D_1$ under the condition \(l = 1\) and \(d = 1\) with varying $n_q$.
We first compared differences in accuracy across insertion layers.
As shown in Figure \ref{fig:rmse-nq}, the RMSE tends to be higher for quantum circuits with \(P = R_{y}\).
This is mainly because of the limited expressivity of the parameterized unitary in the \(R_{y}\) circuit.
In contrast to the \(R_{y}\) circuit, \(R_{z}R_{y}\) circuit yielded more accurate results than the classical NN layer,
even though the number of parameters in the \(R_{z}R_{y}\) circuit was smaller than in the classical NN.
For example, the RMSE of the \(R_{z}R_{y}\) circuits for \(n_{q} = 4\) was 1.48 kcal/mol, and 1.80 kcal/mol for the classical NN.
This tendency was consistent across all tested $n_q$.

However, the improvement in accuracy between the classical NN and the \(R_{z}R_{y}\) circuits became smaller as $n_q$ increased,
indicating that the gain in the classical NN was larger than that in the quantum-circuit-inserted model with respect to $n_q$.
Specifically, RMSEs in the classical NN decreased from 1.80 to 1.40 kcal/mol as $n_q$ increased from \(4\) to \(12\),
whereas those in the \(R_{z}R_{y}\) circuits decreased from 1.48 to 1.27 kcal/mol.
This may be because the number of parameters in the classical NN layer increases with the square of \(n_{q}\).

From a practical viewpoint, the difference between the classical NN and the quantum model is expected to diminish as \(n_{q}\) increases.
Moreover, due to current hardware limitations, increasing \(n_{q}\) in quantum models remains challenging,
whereas classical NNs can readily scale their dimensions to improve accuracy.
Therefore, from a practical perspective, these results do not yet indicate superiority of the quantum approach over classical methods.

\subsubsection{Dependence on $l$}

\begin{figure}
    \centering
    \begin{subfigure}[b]{0.32\textwidth}
        \centering
        \includegraphics[width=\textwidth]{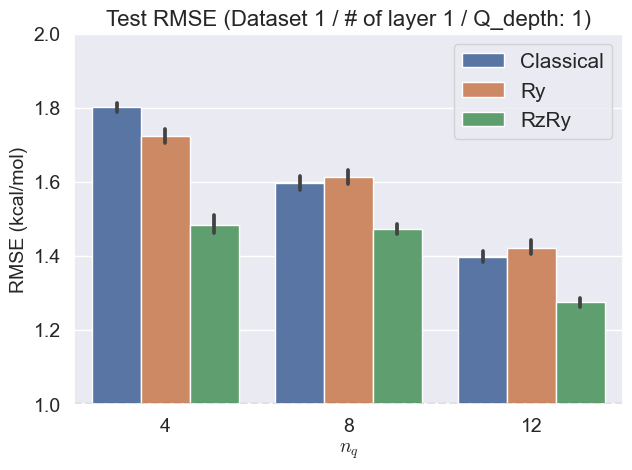}
        \caption{RMSEs vs. $n_q$ with varying circuit types}
        \label{fig:rmse-nq}
    \end{subfigure}
    \hfill
    \begin{subfigure}[b]{0.32\textwidth}
        \centering
        \includegraphics[width=\textwidth]{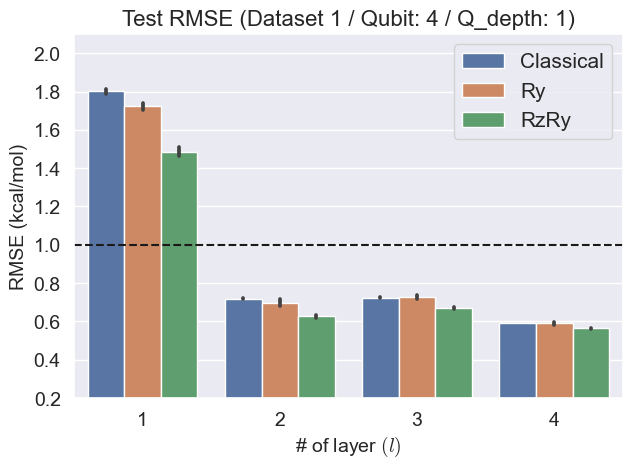}
        \caption{RMSEs vs. $l$ with varying circuit types}
        \label{fig:rmse-l}
    \end{subfigure}
    \hfill
    \begin{subfigure}[b]{0.32\textwidth}
        \centering
        \includegraphics[width=\textwidth]{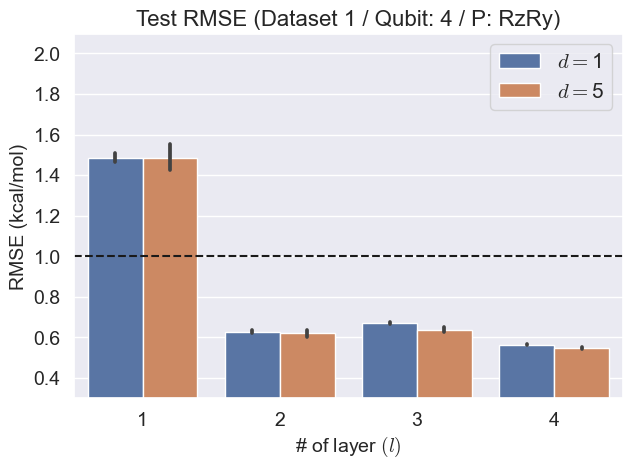}
        \caption{RMSEs vs. $l$ with varying circuit depths}
        \label{fig:rmse-d}
    \end{subfigure}
    \caption{RMSEs (kcal/mol) of transferred models for dataset $D_1$ under different conditions.
    (a) Dependence on the number of qubits, $n_q$, with $l=1$ and $d=1$.
    (b) Dependence on the number of layers, $l$, with $n_q = 4$ and $d=1$.
    (c) Dependence on the quantum-circuit depth, $d$, with $n_q = 4$ and $P = R_z R_y$.
    Error bars represent 95\% confidence intervals calculated from five trials.}
    \label{fig:rmse-summary}
\end{figure}

Next, we examined the RMSE dependence on the number of \(\mathcal{L}_{\text{in}}\) layers, \(l\).
As shown in Figure \ref{fig:rmse-l},
the accuracy difference after transfer learning was largest for \(l = 1\).
This may be because the expressivity of the pretrained model was insufficient, leaving room for improvement through transfer learning;
hence, the effect of $\mathcal{Q}$ appeared more clearly.
Conversely, the difference between the classical NN and quantum models for \(l = 2,\ 3\ \) and \(4\)
tended to be smaller than that for \(l = 1\).
This behavior can be quantified by the improvements over the pretrained
models: \(l = 1,\ \ldots,\ 4\) were 1.55, 0.39, 0.14 and 0.13 kcal/mol, respectively,
for transfer learning with $n_{q} =4$ and \(R_zR_y\) circuits, as shown in Figure \ref{fig:rmse-l}.
This may be because the RMSEs for the test data after pretraining 
were already below the chemical-accuracy threshold of 1 kcal/mol,
 leaving little room for improvement through transfer learning.
These results suggest that ensuring sufficient room for improvement during pretraining is essential
to fully exploit the advantage of the quantum model.

\subsubsection{Dependence on $d$}

\begin{figure}
  \centering
  \begin{subfigure}[b]{0.24\textwidth}
      \centering
      \includegraphics[width=\textwidth]{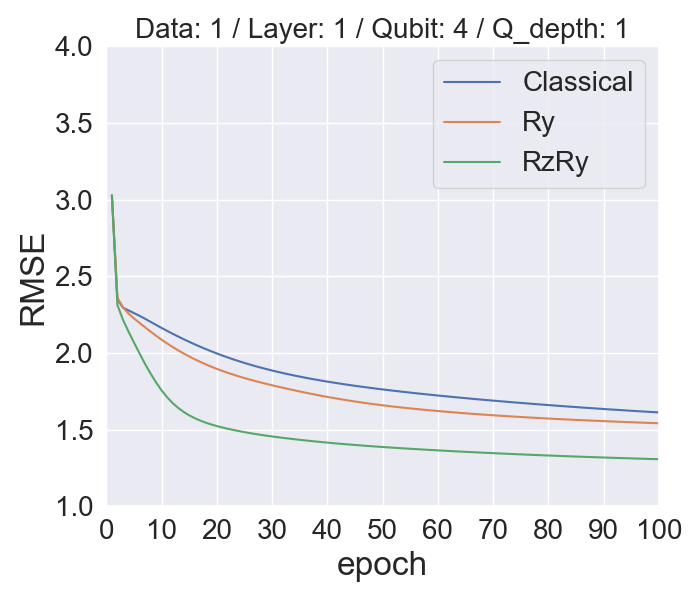}
      \caption{$d=1, l=1$}
  \end{subfigure}
  \hfill
  \begin{subfigure}[b]{0.24\textwidth}
      \centering
      \includegraphics[width=\textwidth]{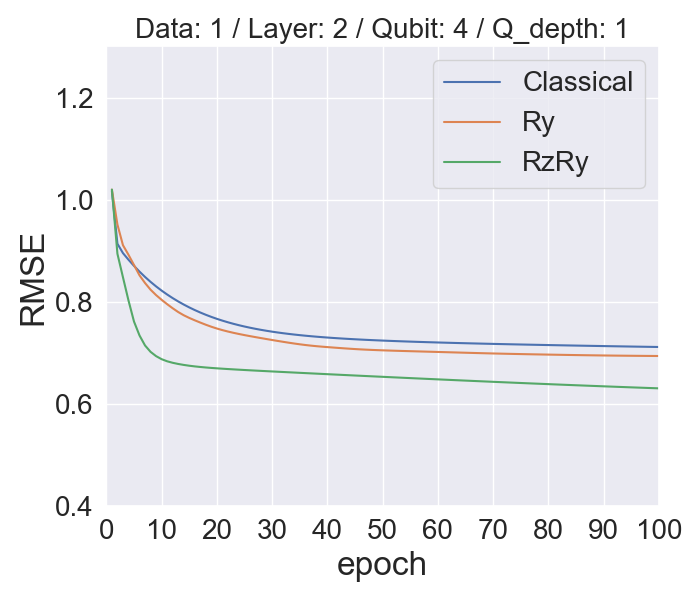}
      \caption{$d=1, l=2$}
  \end{subfigure}
  \hfill
  \begin{subfigure}[b]{0.24\textwidth}
      \centering
      \includegraphics[width=\textwidth]{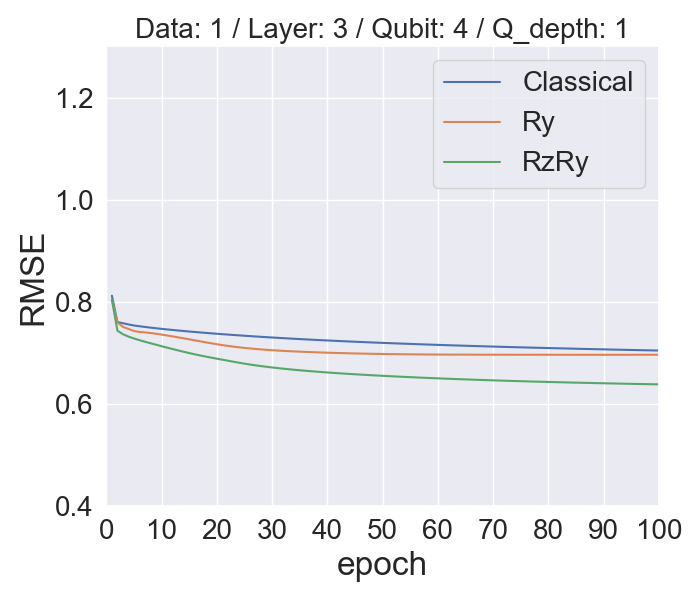}
      \caption{$d=1, l=3$}
  \end{subfigure}
  \begin{subfigure}[b]{0.24\textwidth}
    \centering
    \includegraphics[width=\textwidth]{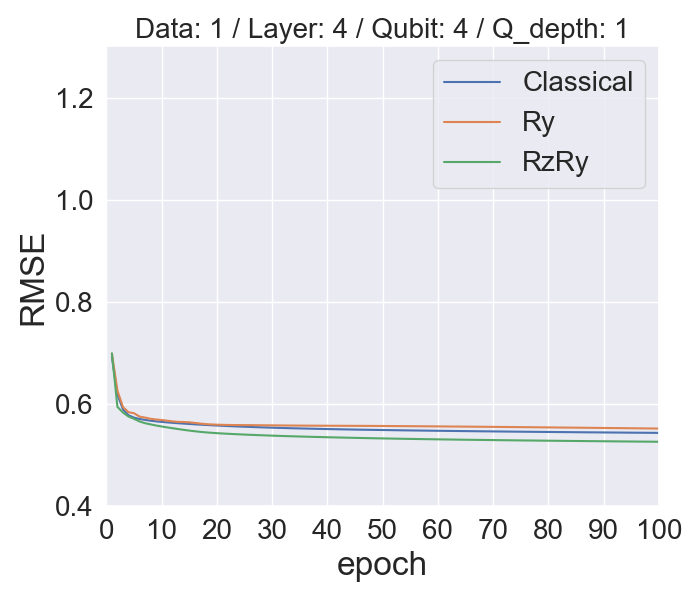}
    \caption{$d=1, l=4$}
  \end{subfigure}
  \\
  \begin{subfigure}[b]{0.24\textwidth}
    \centering
    \includegraphics[width=\textwidth]{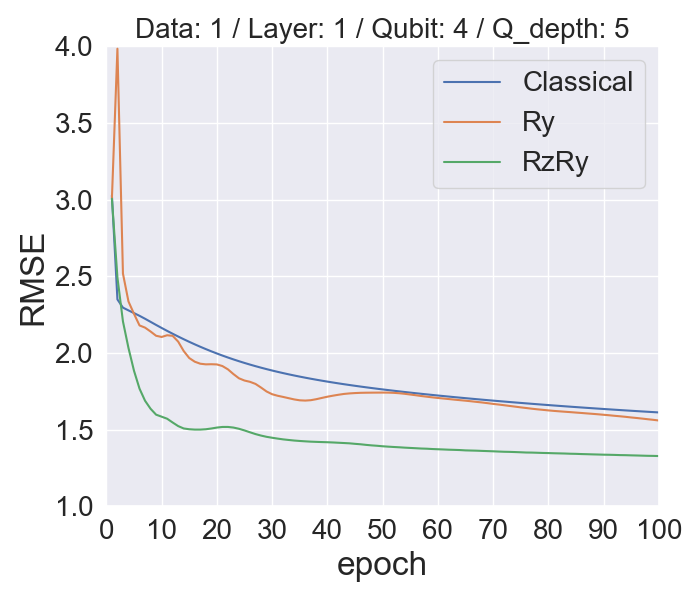}
    \caption{$d=5, l=1$}
  \end{subfigure}
  \hfill
  \begin{subfigure}[b]{0.24\textwidth}
    \centering
    \includegraphics[width=\textwidth]{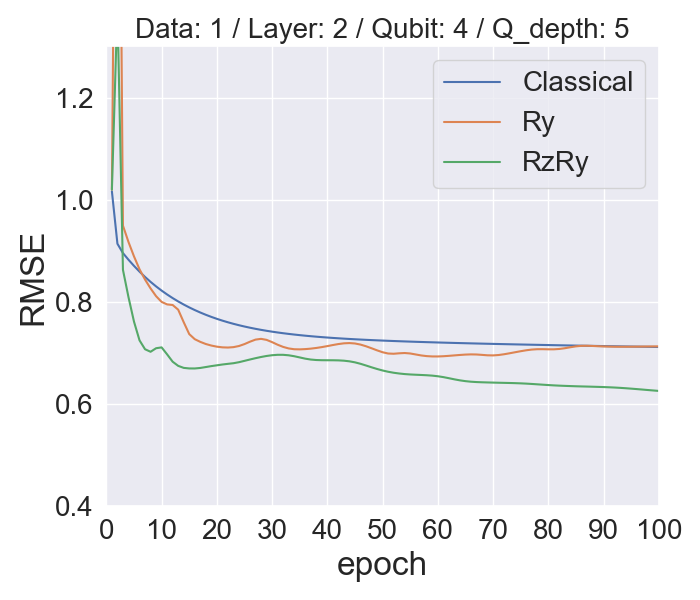}
    \caption{$d=5, l=2$}
  \end{subfigure}
  \hfill
  \begin{subfigure}[b]{0.24\textwidth}
    \centering
    \includegraphics[width=\textwidth]{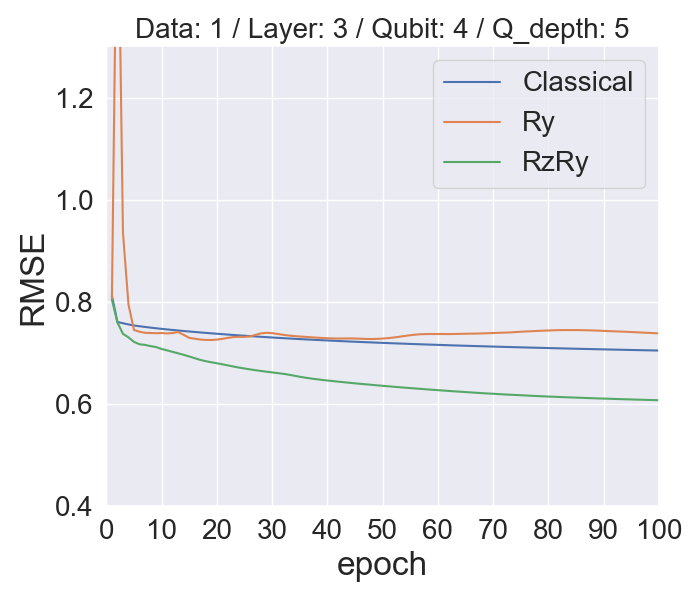}
    \caption{$d=5, l=3$}
  \end{subfigure}
  \begin{subfigure}[b]{0.24\textwidth}
    \centering
    \includegraphics[width=\textwidth]{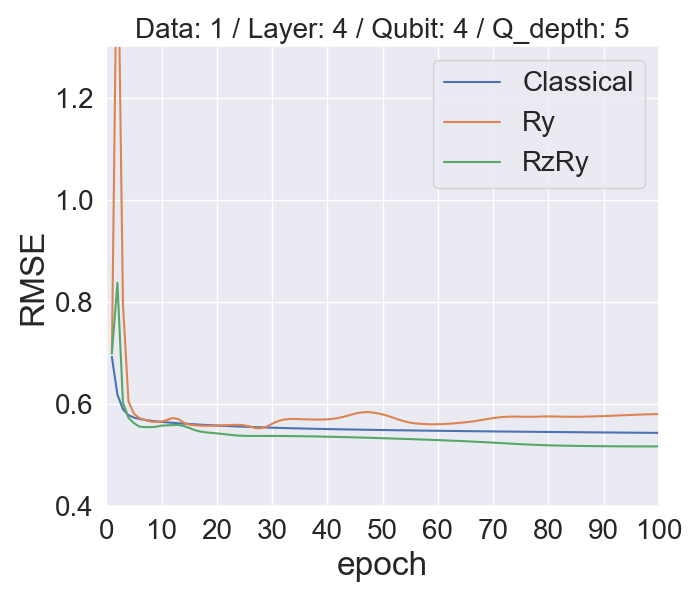}
    \caption{$d=5, l=4$}
  \end{subfigure}
  \caption{Prediction-accuracy transitions for validation data.
  The upper-row panels ((a)--(d)) show the results for $d=1$ with varying $l$ values, and the lower-row panels ((e)--(h)) show those for $d=5$.}
  \label{fig:learning_curve}
\end{figure}

We now discuss the effect of circuit depth \(d\).
Increasing \(d\) from 1 to 5 would typically be expected to enhance the model's expressive power.
However, as shown in Figure \ref{fig:rmse-d}, there were almost no differences in RMSE between \(d = 1\) and \(5\) for all \(l\).
This may be because the deeper $d=5$ circuit is more difficult to optimize effectively.
The results in Figure \ref{fig:learning_curve} indicate that training the quantum-circuit model,
particularly for \(d = 5\), was unstable at the beginning of optimization.
Therefore, it may be necessary to reconsider the optimization strategy, including the learning rate,
number of epochs, and other training hyperparameters.
To mitigate this issue, one possible approach is to incrementally add and optimize quantum-circuit layers one at a time,
thereby improving training stability and efficiency, as suggested in Ref. \cite{Skolik_2021}.

\subsubsection{Dataset dependence}
\begin{figure}
  \centering
  \includegraphics[width=3.93701in,height=2.92775in]{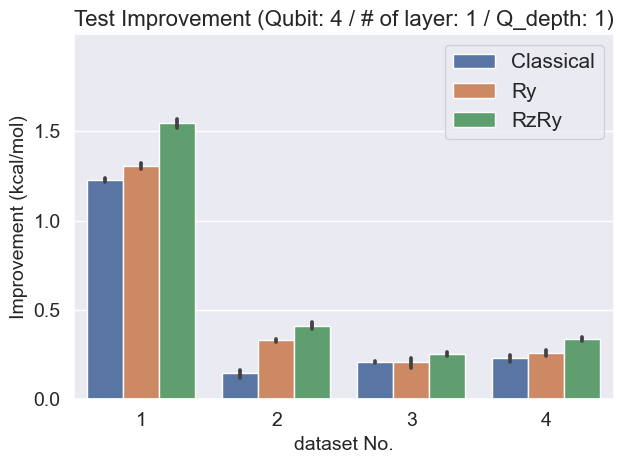}
  \caption{Improvements (kcal/mol) of transferred models on the test data
    for datasets \(D_1\) to \(D_4\) under \(n_{q} = 4\), \(l = 1\) and \(d = 1\).
    The vertical axis represents improvements (kcal/mol) from
    the pretrained models, and the horizontal axis represents the dataset number.
    Error bars indicate 95\% confidence intervals calculated from five trials.}
  \label{fig:improvement-dataset}
\end{figure}

Finally, we examine differences among the four datasets.
The improvements relative to the pretrained model are summarized in terms of RMSE in Figure \ref{fig:improvement-dataset}.
Relatively small differences between the
quantum and classical models were observed for dataset \(D_3\) and \(D_4\) compared with \(D_1\).
This likely results from differences in pretraining progress.
The number of samples used for pretraining in datasets \(D_3\) and \(D_4\)
was much larger than that in \(D_1\) (see Table \ref{table:dataset}).
We believe that this achieved higher initial accuracy, leaving little room for further improvement during transfer learning.

\subsection{Numerical assessment on practical molecule: cholesterol and its isomers}\label{results_cholesterol}
\begin{figure}
  \centering
  \begin{subfigure}[b]{0.24\textwidth}
      \centering
      \includegraphics[width=\textwidth]{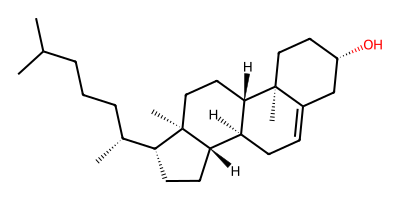}
      \caption{Cholesterol}
  \end{subfigure}
  \hfill
  \begin{subfigure}[b]{0.24\textwidth}
      \centering
      \includegraphics[width=\textwidth]{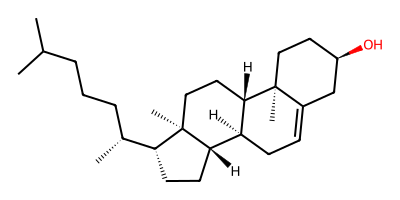}
      \caption{Epicholesterol}
  \end{subfigure}
  \hfill
  \begin{subfigure}[b]{0.24\textwidth}
      \centering
      \includegraphics[width=\textwidth]{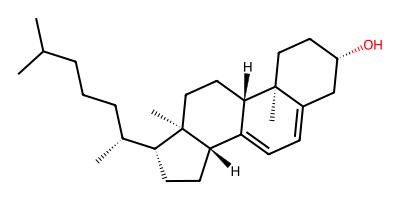}
      \caption{7-Dehydrocholesterol}
  \end{subfigure}
  \begin{subfigure}[b]{0.24\textwidth}
      \centering
      \includegraphics[width=\textwidth]{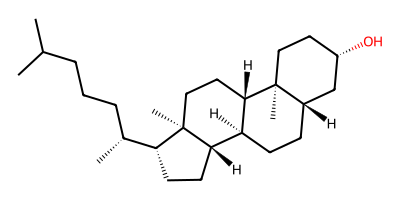}
      \caption{Cholestanol}
  \end{subfigure}
  \caption{Skeletal structures of cholesterol and three closely related sterol isomers
  used as a practical validation set for assessing transferability to larger molecules.}
  \label{fig:valid_molecules}
\end{figure}

In Sec. \ref{results_ani}, our models were validated using the ANI dataset, mainly $D_1$, 
which includes simple molecules such as CH\textsubscript{4}, NH\textsubscript{3} and H\textsubscript{2}O.
Here, we apply our models to cholesterol and its isomers to evaluate performance on more complex,
realistic molecular systems.
The molecular structures are shown in Figure \ref{fig:valid_molecules}.

For accuracy verification, DFT calculations were performed for each target molecule,
and structural optimization was carried out at the B3LYP/6-31G(d) level on Gaussian 16 \cite{g16}.
Using the optimized atomic coordinates as input, molecular energies were
estimated with the original ANI, the classical-transfer, and the quantum-transfer models.
For transfer learning, dataset $D_4$ was used as training data, and $R_z R_y$ circuits were employed in the quantum-transfer model.
The predicted energies were then compared with the DFT-calculated energies for each molecule.
Hyperparameters were set as $d=1$ and $n_q=4$, and $l=1, 2$ or $3$.

\begin{figure}
  \centering
  \begin{subfigure}[b]{0.24\textwidth}
      \centering
      \includegraphics[width=\textwidth]{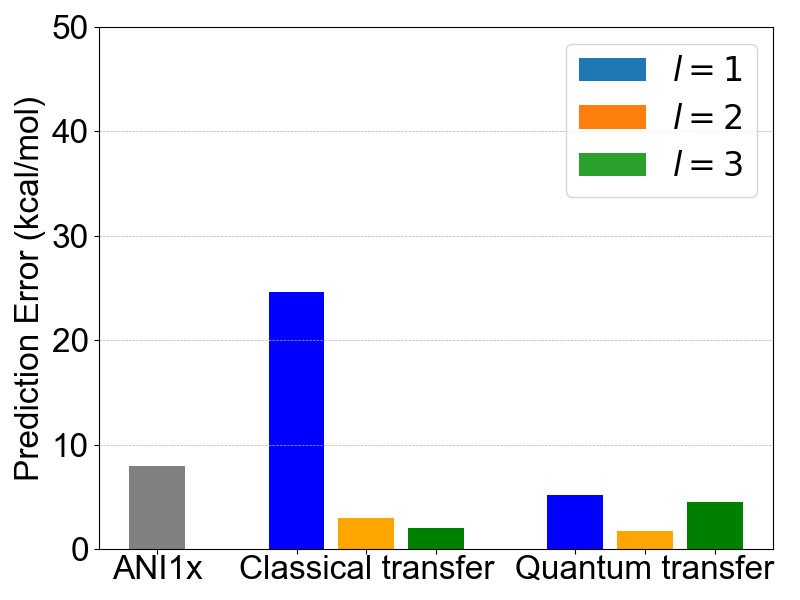}
      \caption{Cholesterol}
  \end{subfigure}
  \hfill
  \begin{subfigure}[b]{0.24\textwidth}
      \centering
      \includegraphics[width=\textwidth]{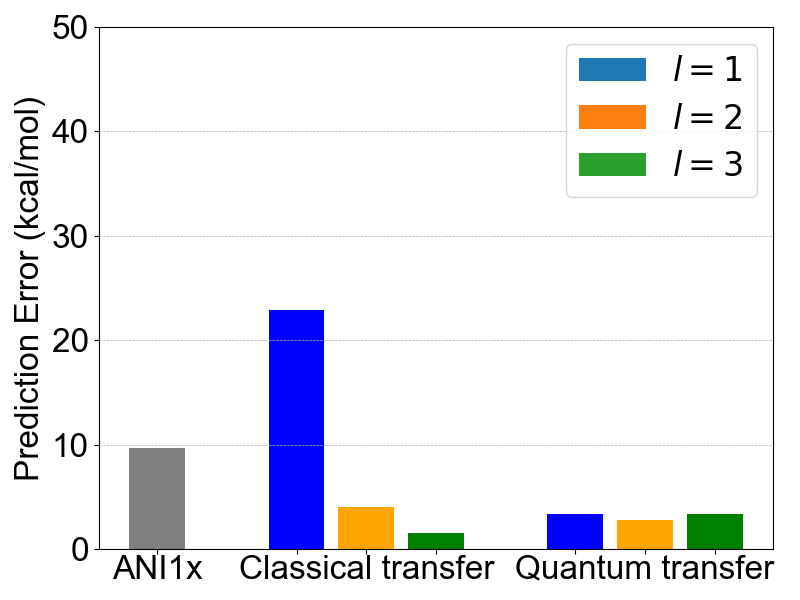}
      \caption{Epicholesterol}
  \end{subfigure}
  \hfill
  \begin{subfigure}[b]{0.24\textwidth}
      \centering
      \includegraphics[width=\textwidth]{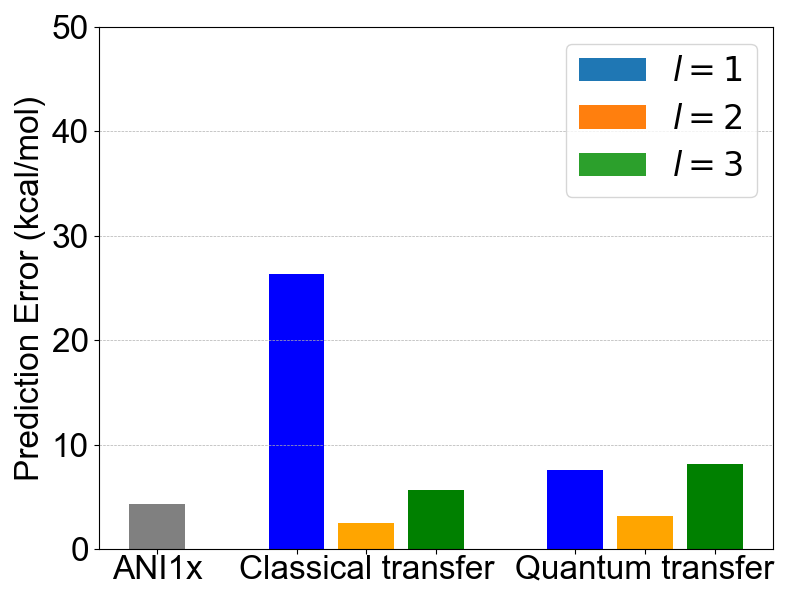}
      \caption{7-Dehydrocholesterol}
  \end{subfigure}
  \begin{subfigure}[b]{0.24\textwidth}
    \centering
    \includegraphics[width=\textwidth]{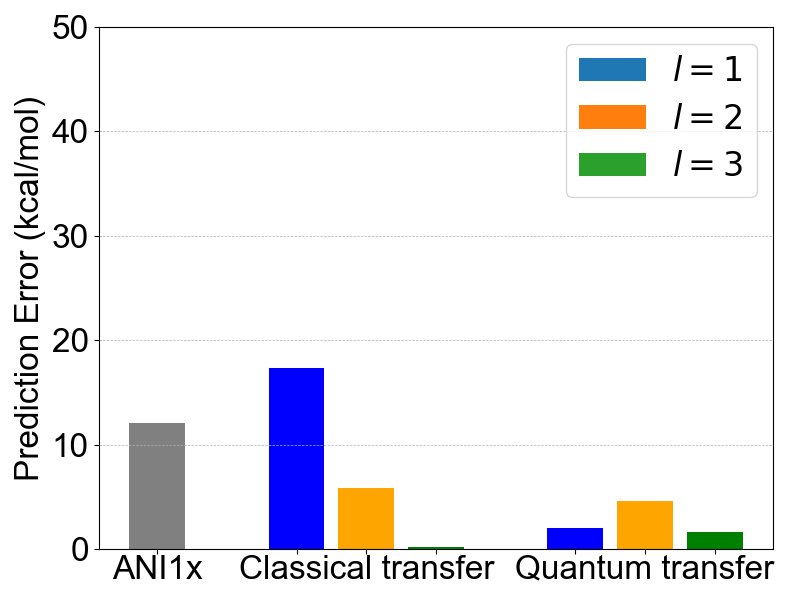}
    \caption{Cholestanol}
  \end{subfigure}
  \caption{Prediction errors (kcal/mol) from DFT-calculated values for cholesterol and its isomers.
  For transfer models, the number of \(\mathcal{L}_{\text{in}}\) layers, \(l\) values were varied from 1--3.
  Except for 7-Dehydrocholesterol, quantum transfer models outperformed the original ANI model for all $l$ values.}
  \label{fig:cholesterol_result}
\end{figure}

Figure \ref{fig:cholesterol_result} shows the absolute errors between the predictions of each model and the DFT-calculated values.
As illustrated, the quantum-transfer model predicts molecular energies with high accuracy,
showing errors smaller than 10 kcal/mol relative to DFT for all molecules.
Except for 7-Dehydrocholesterol, it outperformed the original ANI model,
confirming that the quantum model remains effective even for larger molecular systems.
For 7-Dehydrocholesterol, which contains a sequence of conjugated carbon-carbon double bonds,
accuracy did not improve, likely because such molecular motifs are rare in the transfer-learning dataset $D_4$.

In the DFT calculations, the energy ordering was Cholestanol $<$ Cholesterol $<$ Epicholesterol $<$ 7-Dehydrocholesterol.
All quantum models successfully reproduced this energetic order among the structural isomers,
suggesting that the proposed method can be applied to evaluate the relative stability of structural isomers.

Overall, the proposed hybrid approach achieved accuracy comparable to—or,
depending on the molecule and training configuration,
higher than—that of the original machine-learning potential when applied to realistic molecular systems.

\section{Conclusion}\label{conclusion}
In this study, we proposed and evaluated a quantum-classical hybrid framework
for constructing MLIP.
Specifically, we incorporated QCL into the ANI neural-network architecture to enhance molecular-energy prediction accuracy.
The central concept was to replace the final layer of the ANI model with a QCL module,
leveraging the pretrained classical neural network to reduce the dimensionality and complexity of the quantum feature space.
Our numerical experiments evaluated this hybrid approach across various settings,
including different numbers of qubits ($n_q$), circuit depths ($d$), pretrained network layers ($l$) and datasets.

Our results showed that the QCL-enhanced model outperformed the purely classical neural network under specific conditions—particularly
for smaller molecular datasets and shallower pretrained models (small $l$)
where the classical model retained capacity for improvement.
The quantum model employing an $R_z R_y$ circuit with $n_q = 4$ demonstrated
the most notable enhancement, achieving a lower RMSE compared with a conventional neural-network layer.
Validation using cholesterol and its isomeric molecules further confirmed that,
under certain learning configurations, the hybrid model achieved accuracy comparable to or exceeding that of the original ANI potential
for realistic molecular systems.

We also observed that benefit of quantum transfer learning depended strongly on the state of the pretrained model.
When the pretrained classical network had already achieved high accuracy, the QCL model contributed only marginal improvements.
This indicates that quantum-enhanced MLIPs are particularly advantageous when
the classical model is limited by insufficient data or network capacity, leaving representational gaps that quantum features can address.
In practice, however, such scenarios are uncommon because classical networks can often enhance accuracy by expanding their dimensionality,
which makes it challenging to demonstrate a clear quantum advantage with current methods.

Overall, this work provides insights into the feasibility and limitations of integrating quantum neural networks into MLIPs.
While QCL-based models yielded promising results under selected conditions,
further progress will require improving circuit expressivity, optimizing training algorithms, and scaling to larger and more chemically diverse datasets.
Future investigations may also explore experimental deployment on real quantum hardware to evaluate robustness against noise and hardware-specific constraints.
By overcoming these challenges, quantum-enhanced MLIPs could contribute to more accurate and computationally efficient molecular simulations,
thereby advancing the broader fields of quantum chemistry and materials informatics.

\section{Acknowledgments}\label{acknowledgments}
This work was supported by MEXT Quantum Leap Flagship Program (MEXT Q-LEAP) 
Grant No. JPMXS0120319794, the JST COI-NEXT Program Grant No. JPMJPF2014, 
and the JST ASPIRE Program Grant No. JPMJAP2319. 

\bibliography{bibliography}
\bibliographystyle{unsrt}

\end{document}